\documentstyle[prl,aps,epsfig,twocolumn]{revtex}
\begin{document}

\title{Laser cooling all the way down to molecular condensate}

\author{J. Dziarmaga$^1$ and  M. Lewenstein$^2$}

\address{
$^1$ 
Institute of Physics and Centre for Complex Systems,
Jagiellonian University,
ul.~Reymonta 4, 30-059 Krak\'ow, Poland \\
$^2$
Institut f\"ur Theoretische Physik,
Universit\"at Hannover, 
D-30167 Hannover, Germany
}

\date{ January 25, 2005 }
\maketitle

\begin{abstract}
Numerical simulations show that laser cooling of fermions on the repulsive
side of the Feshbach resonance can sympathetically cool molecules
well below their condensation temperature.
\end{abstract}


Fermi superfluids are in the center of interest of recent studies in 
physics of ultracold quantum gases.  They provide perhaps the most 
promising model system to study
the  superfluidity, and  the Bardeen-Cooper-Schrieffer (BCS) theory
in the limit of strong interactions, when the size of Cooper pairs
becomes comparable or even smaller than average interparticle distance,
and the mean field description breaks down \cite{Randeria}.
In cold atom systems  the use of 
Feshbach resonances permits to tune the $s$-wave scattering length 
between the
atoms with opposite ``spin'' \cite{Feshbach}, assuring in this way a 
  high flexibility of the systems. This technique allows in particular to 
drive the system through the BCS-BEC crossover when the scattering length
diverges and changes sign from negative (attractive) to positive 
(repulsive).
During this passage,  the state of the system continuously evolves from 
the BCS
superfluid of weakly correlated Cooper pairs, through the pseudogap 
regime,
and then the unitarity limit, when the scattering length diverges,  
to the Bose-Einstein condensate (BEC) of diatomic 
molecules.
By now several successful experiments with molecular condensates were 
performed
\cite{mBEC} which opened a way to first successful experiments with Fermi
superfluids \cite{Grimm}.

Earlier
theoretical studies demonstrated that laser cooling allows to reach
the degeneracy in Fermi systems \cite{Idziaszek_f}, and  
 to go all the way down to the Fermi
superfluid \cite{BCS} on the attractive
side of the Feshbach resonance.
The purpose of this Communication is to show that laser cooling offers 
also an alternative technique
to reach molecular condensate on the repulsive side of the resonance.
Moreover, it  allows  to post-cool the condensate prepared via
evaporative cooling down to even lower temperatures. We have developed 
here 
a fully self-consistent quantum kinetic theory
of molecular condensate creation via laser cooling \footnote{similar to
self-consistent theories of the BEC growth \cite{GZ}}. The results are 
very
promising: despite the fact that condensate formation shifts the energies
of quasiparticle excitations with respect to the equally spaced harmonic
oscillator levels, and at the same time reduces fermionic population 
exposed
to laser cooling, the cooling remains efficient well below the 
condensation
temperature of $0.18 T_F$ down to at least $0.06T_F$. Optimization of the
cooling protocol could probably significantly reduce this temperature.


We consider system containing two species of fermions with ``spin up
and down'', molecules which are in their bound states, and fermions in
an excited state. We work on the repulsive side of the Feshbach resonance,
where interactions strengths between spin up/down fermions and molecules,
as well as the binding energy of the molecules are simple universal
functions of the effective $s$-wave scattering length $a$ between spin
up and down fermions \cite{Petrov}. Close to the resonance the binding
energy of molecules $\nu=-\frac{\hbar^2}{ma^2}$ can be made less 
than the condensation
temperature $k_BT_C$ for the molecules. As a result at all temperatures
greater than $T_C$ populations of molecules and fermions are comparable.
Below $T_C$, a fraction of molecules condenses leaving room in molecular
phase space for more fermions to bind into molecules. Fermionic population 
is then shrinking, but it 
remains still comparable to the shrinking population of non-condensed 
molecules
all the way down to the dissociation temperature, $T_D=\nu/k_B$. 
Below $T_D$, fermionic population is suppressed,
but at the same time almost all molecules are already condensed. Thus,
close to the Feshbach resonance, when $T_D<T_C$,  there are favorable
conditions for efficient sympathetic cooling of molecules by collisions
with spin up/down fermions at all temperatures of interest. The fermions 
in turn are subject to laser cooling, and provide 
the heat reservoir, which is comparable to the cooled system
(non-condensed molecules). Populations of the spin up and down fermions 
are
the same. Mutual $s$-wave interactions between these two populations and 
the
population of molecules lead to thermalization on a time scale,
 which is faster
than the rate of laser cooling. At the same time three-body collisions 
keep
the system close to chemical equilibrium between molecules and fermions. 
The
system remains thus in a state of quasi-equilibrium with slowly decreasing
temperature. This thermal state is described by the generalized
BCS theory or boson-fermion model \cite{BF} which includes BEC 
 of molecules described by the Gross-Pitaevskii equation.
The model takes into account coherent tunneling between molecules and 
pairs
of spin up and down fermions. Thermal excitations of the molecular
condensate are described by the bosonic Bogoliubov theory. Structure of
the ground state of fermions and its fermionic excitations follows from 
the
set of Bogoliubov-de Gennes equations together with self-consistent
definitions of the gap function and the mean field potential. On the
repulsive side of the Feshbach resonance the leading contribution to the
gap function comes from the molecular BEC through the coherent
coupling between molecules and pairs of fermions. In particular the gap
function experienced by fermions is proportional to the condensate
wave function. Laser cooling of fermions was described in detail in
Ref.\cite{Idziaszek_f}. Coherent laser excites atoms from, say, the
spin down ground state to the excited state, and spontaneous
emission brings them back to the spin down ground state. Frequencies of 
the
laser assure that the generalized Raman cooling takes place
\cite{Idziaszek_b}. In this Communication we take into account 
interactions
between fermions and describe the cooling process in terms of 
instantaneous
fermionic and bosonic (molecular) Bogoliubov quasiparticles, whose
eigenfunctions and eigenenergies are self-consistently updated during the
evolution. We work in the ``festina lente'' limit to avoid
reabsorbtion effects \cite{FL}, and also employ spherical symmetry and
ergodic approximations \cite{BCS,GZ}.

The boson-fermion model \cite{BF} is defined by the Hamiltonian
\begin{eqnarray}
\hat H=
&\int d^3r&
\left[
\sum_{a=+,-,e,m}
\hat\psi^{\dagger}_a
{\cal H}_a
\hat\psi_a +
(\nu-\mu)
\hat\psi^{\dagger}_m\hat\psi_m +
\right.
\nonumber\\
&&
\lambda
\hat\psi^{\dagger}_m
\hat\psi_{+}
\hat\psi_{-}+   
{\rm ~h.c.}+
\nonumber\\
&&
g
\hat\psi_{+}^{\dagger}
\hat\psi_{-}^{\dagger} 
\hat\psi_{-}
\hat\psi_{+}+
0.3~g
\hat\psi_m^\dagger
\hat\psi_m^\dagger
\hat\psi_m
\hat\psi_m+
\nonumber\\
&&
\left.
1.2~g
\hat\psi_m^\dagger
\hat\psi_m
\left(
\hat\psi_{+}^{\dagger}
\hat\psi_{+}+
\hat\psi_{-}^{\dagger}
\hat\psi_{-}
\right)
\right]~,
\label{HBF}  
\end{eqnarray}
where the fields $\hat\psi_{+}$ and $\hat\psi_{-}$ describe fermions with 
spin up
and down, respectively, $\hat\psi_e$ corresponds to fermions in the 
excited
state, and $\hat\psi_m$ is the bosonic molecular field. Here
${\cal H}_a=-\frac{\hbar^2}{2m_a}\nabla^2+V_a(\vec r)-\mu$ is a
single particle Hamiltonian, $m_a$ is atomic mass $m$ or molecular mass 
$2m$,
$V_a(\vec r)=\frac12m_a\Omega_a^2r^2$ is the harmonic trap potential,
$\mu$ is chemical potential, $g=\frac{4\pi\hbar^2 a}{m}$ is 
fermion-fermion
interaction strength with a large positive $s$-wave scattering length $a$,
$\lambda$ is a coupling between pairs of atoms and molecules, and
$\nu=-\frac{\hbar^2}{ma^2}$ is the molecular binding energy. Close to
the Feshbach resonance, when $g$ is large, we can neglect 
interactions with excited atoms.

In the mean field approximation, which closely follows the BCS theory,
the quartic fermion-fermion interaction term in the Hamiltonian 
(\ref{HBF})
is made quadratic by replacing products of operators by their averages
in all possible ways. The averages are the mean field potential
$W(\vec r)=
1.2~g
\left\langle
\hat\psi^{\dagger}_m(\vec r)
\hat\psi_m(\vec r)
\right\rangle+
g
\left\langle
\hat\psi^{\dagger}_{\pm}(\vec r)
\hat\psi_{\pm}(\vec r)
\right\rangle $,
and the anomalous pairing potential 
$P(\vec r)=
\left\langle
\hat\psi_+(\vec r)
\hat\psi_-(\vec r)
\right\rangle $,
which is mixing fermions with spin up and down. In the mean-field 
approximation
the molecular field  $\hat\psi_m$ is replaced by a $c$-number
condensate amplitude $\phi$ that fulfills the stationary
Gross-Pitaevskii equation
\begin{equation}
(2\mu-\nu)\phi={\cal H}_m\phi+0.6~g|\phi|^2\phi+\lambda P~.
\label{GPE}
\end{equation}
Here $\phi$ is normalized to the number of condensed molecules. The
mean-field equations for fermions are
\begin{eqnarray}
&&
i\hbar\frac{d}{dt}~\hat\psi_{\pm}~=~
{\cal H}_\pm~\hat\psi_{\pm}+
W(\vec r)~\hat\psi_{\pm}~\mp~
\Delta(\vec r)~
\hat\psi^{\dagger}_{\mp}~,
\label{pm2}
\end{eqnarray}
with an effective gap function
$\Delta(\vec r)=gP(\vec r)+\lambda\phi(\vec r)~.$
These equations are mixing spin up and down components but
they can be ``diagonalized'' by the Bogoliubov transformation
$\hat\psi_\pm(\vec r)~=~
\sum_m~
\hat b_{m,\pm}~           u_m(\vec r) ~\mp~
\hat b_{m,\mp}^{\dagger}~ v_m^*(\vec r)$, 
with fermionic quasiparticle annihilation operators $\hat b_{m,\pm}$.
The quasiparticle  modes $(u_m,v_m)$ fulfill the Bogoliubov-de Gennes
equations with positive energies  $\omega_m$,
\begin{eqnarray}
&&
\omega_m~u_m~=~ 
+~{\cal H}_\pm~u_m~+~
W~u_m~
-~\Delta~v_m~,
\nonumber\\
&&
\omega_m~v_m~=~ 
-~{\cal H}_\pm~v_m~-
~W~v_m~
-~\Delta^*~u_m ~.
\label{BdG}
\end{eqnarray}
In a
thermal state with inverse temperature $\beta$ the average occupation
numbers of quasiparticle states are
$N_m=\langle \hat b^{\dagger}_{m,\pm} \hat b_{m,\pm} \rangle=
[\exp(\beta\omega_m )+1]^{-1}$. Equations (\ref{BdG}) are solved
together with the self-consistency conditions
\begin{eqnarray}
\Delta(\vec r)&=&
\lambda \phi(\vec r)+
g
\sum_m
\left[ 1 - 2 N_m \right] u_m(\vec r) v_m^*(\vec r) ~,  
\label{Puv}\\
W(\vec r)&=&
1.2~g
\langle
\hat\psi_m^\dagger(\vec r) \hat\psi_m(\vec r)
\rangle +
\nonumber\\
&&
g
\sum_m
(1-N_m) |v_m(\vec r)|^2 +
N_m     |u_m(\vec r)|^2~,
\label{Wuv}
\end{eqnarray}
by successive iterations with the chemical potential $\mu$ adjusted to 
keep
 the total number of atoms constant.
The average molecular density
$\langle\hat\psi_m^\dagger(\vec r) \hat\psi_m(\vec r)\rangle$ is a sum of
the condensate density $|\phi(\vec r)|^2$ plus density of non-condensed
molecules. After every iteration the Gross-Pitaevskii equation (\ref{GPE})
is solved by relaxation. The total number of atoms includes atoms bound 
into
non-condensed molecules depleted from the condensate by thermal and
quantum fluctuations. We estimate the number of non-condensed molecules
using the approximate bosonic Bogoliubov modes in the Thomas-Fermi
limit \cite{Stringari}. The ultraviolet divergence in Eq.(\ref{Puv})
is regularized by the quickly convergent method of Ref.\cite{LDA}.

Excitation of atoms from the spin down state to the excited state is
described by the Hamiltonian
\begin{equation}
\hat H_{\rm las}=
\int d^3r~
\left[
\frac12\Omega
e^{i\vec k_{L}\vec r}
\hat\psi_{e}^{\dagger}(\vec r)
\hat\psi_{-}(\vec r)+
{\rm h.c.}-
\delta
\hat\psi^{\dagger}_e
\hat\psi_e
\right],
\end{equation}
driving coherent oscillations with Rabi frequency $\Omega$ and laser
detuning $\delta$. The excitation is accompanied by the spontaneous
emission described by a superoperator
\begin{eqnarray}
{\cal L}~\hat\rho&=&
\gamma 
\sum_{mk} 
U_{mk}
{\cal D}[\hat b_{m,-}^{\dagger}\hat e_k]\rho +
V_{mk}
{\cal D}[\hat b_{m,+}\hat e_k]\rho
\end{eqnarray}
with a spontaneous emission rate $\gamma$. Here, the Lindblad
superoperator is
${\cal D}[\hat A]\rho=\hat A\rho\hat A^{\dagger}-
\frac12\hat A^{\dagger}\hat A-\frac12\hat A^{\dagger}\hat A\rho$, and
the matrix elements are e.g.
$
U_{mk}=
\int d\Omega_k  {\cal W}(\Omega_k)
|u_{mk}(\vec k)|^2
$
with the spontaneous emission pattern ${\cal W}(\Omega_k)$ and the
generalized Frank-Condon factors
$
u_{mk}(\vec k)=
\int d^3r~
e^{i\vec k \vec r}
w_k^*(\vec r)
u_m(\vec r)
$.
Here, $w_k(\vec r)$ is the $k$-th eigenstate of the harmonic 
oscillator and $\hat e_k$ is an annihilation operator of an excited
atom in this state. 

Adiabatic elimination of the excited state (cf.
\cite{AE}) leads to the kinetic equations for the occupation
numbers taking into account the effects of Fermi statistics
\begin{eqnarray}
\frac{d N_{m,-}}{dt} &=&
\sum_n~
\Gamma^{(-)}_{m\leftarrow n} 
\left( 1 - N_{m,-} \right) N_{n,-} -
(n\leftrightarrow m)+
\nonumber\\
&&
C_{mn} \left( 1 - N_{m,-} \right) \left( 1 - N_{n,+} \right) -
A_{nm} N_{m,-} N_{n,+} ~,
\nonumber\\
\frac{d N_{m,+}}{dt} &=&
\sum_n~
\Gamma^{(+)}_{m\leftarrow n} 
\left( 1 - N_{m,+} \right) N_{n,+} -
(n\leftrightarrow m)+
\nonumber\\
&&
C_{nm} \left( 1 - N_{m,+} \right) \left( 1 - N_{n,-} \right) -
A_{mn} N_{m,+} N_{n,-} ~.
\nonumber
\end{eqnarray}
They describe relaxation of $\pm$ quasiparticles
$(\Gamma^{(\pm)}_{m\leftarrow n})$, and creation/annihilation of pairs of
$+$ and $-$ quasiparticles $(C_{nm}~{\rm and }~A_{nm})$. The transition
rates are e.g.
\begin{equation}
\Gamma^{(-)}_{m\leftarrow n}=
\frac{\Omega^2}{2\gamma}
\sum_k
\frac{
\gamma^2 
U_{mk} 
|u_{nk}(\vec k_L)|^2
}
{
(\delta-\omega^e_k+\omega_n)^2+
\gamma_k^2
}~.
\label{Gamma-}
\end{equation}
Here, $\omega^e_k$ is the energy of the $k$-th harmonic oscillator
state and $\gamma_k$ is approximate spontaneous decay rate of an
excited atom in this state,
$\gamma_k=\gamma\sum_mU_{mk}(1-N_{m,-})+V_{mk}N_{m,+}$, see Appendix A.

Laser cooling drives average occupation numbers $N_{m,\pm}$ of
fermionic quasiparticles out of thermal equilibrium. At the
same time interactions between fermionic and bosonic quasiparticles
drive the system towards thermal equilibrium. Numerical simulations in
Ref.\cite{Idziaszek_f} show that thermal relaxation remains efficient
all the way down to the temperature of $0.03~T_{\rm F}$. This justifies
our assumption of fast equilibration to a quasi-equilibrium state. After
a short period of laser cooling $dt$ the occupation numbers $N_{m,\pm}$ go
out of equilibrium where $N_{m,\pm}=N_{m}=[exp(\beta\omega_m)+1]^{-1}$. 
The initial
total energy $E(T)~=~\sum_{m,\pm}\omega_m N_{m,\pm}+
\sum_m\omega^B_m N^B_m$, including bosonic Bogoliubov quasiparticles with
frequencies $\omega^B_m$, and average occupation
numbers $N^B_m=[\exp(\beta\omega^B_m)-1]^{-1}$, changes by
$dE~=~\sum_{m} \omega_m~(N_{m,+}+N_{m,-}-2N_m)$. The relaxation
brings the occupation numbers $N_{m,\pm}$ to a new state of
equilibrium at a temperature $T+dT$, but does not change the total
energy. The energy of the system at the new temperature $E(T+dT)$ differs
from the initial $E(T)$ by $dE_{T}\approx
2\frac{dT}{T^2}\sum_m\omega_m^2 N_m(1-N_m)+
\frac{dT}{T^2}\sum_m(\omega^B_m)^2 N^B_m(1+N^B_m)$. Conservation of energy
in thermal relaxation means that $dE=dE_{T}$. In our simulations we use
this equality to find the new lower temperature $T+dT$, and then solve
the Bogoliubov-de Gennes equations and the Gross-Pitaevskii equation
self-consistently to adjust the Bogoliubov modes, the condensate wave 
function
and the chemical potential to the new lower temperature. With the new
Bogoliubov modes we calculate new transition rates, etc.

Calculation of the transition rates is the most time consuming part
of the numerical simulation. This is why we were forced to assume
spherical symmetry. For spherically symmetric $\Delta(r)$, $W(r)$
and $\phi(r)$ the Bogoliubov modes are
$u_{ln}(r)~Y_{lm}(\theta,\varphi)$ and their energies $\omega_{ln}$
do not depend on $m$. The bosonic quasiparticle energies in the
Thomas-Fermi limit \cite{Stringari} become
$\omega^B_{ln}=\Omega_m(2n^2+2nl+3n+l)^{1/2}$.
To be consistent with the spherical symmetry we also made ergodic
approximation that quasiparticle occupation numbers $N_{ln,\pm}$ do not
depend on $m$. In other words, we assume fast equilibration within
each quasiparticle energy shell which keeps the system in a spherically
symmetric state. More on spherical symmetry in Appendix B.

In our simulations we use 81 harmonic oscillator levels, and assume
that there are $N=10660$ atoms with spin up and down which in the
non-interacting case at $T=0$ is equivalent to 39 filled energy
levels. The frequency of the isotropic trap is $\Omega_a=2\pi~2400$Hz,
the same for all fermions and molecules. The scattering length for the
interaction between the two species is  $a=1200$\AA, a realistic
value for interactions between two spin states $|F=9/2,m_F=9/2\rangle$
and $|F=9/2,m_F=7/2\rangle$ of $^{40}$K near the Feshbach resonance. In
the natural harmonic oscillator units this scattering length gives the
interaction strength $g=4.7$, and the molecular binding energy $\nu=-7.3$.
The wavelength of the cooling laser is $\lambda=720$nm, and the laser 
detuning
is $\delta=-12\hbar\Omega_a$. In all simulations the frequency of Rabi
oscillations is much less than the spontaneous emission rate,
$\Omega=0.1\gamma$, so that the average occupation of excited state 
remains
small.  In our simulations we neglect $W(r)$. The main effect of
the mean field potential is an overall shift of energy levels which can
be compensated by laser detuning $\delta$. The coherent tunneling
rate in trap units is $\lambda=1$.

\begin{figure}
\centering
{\epsfig{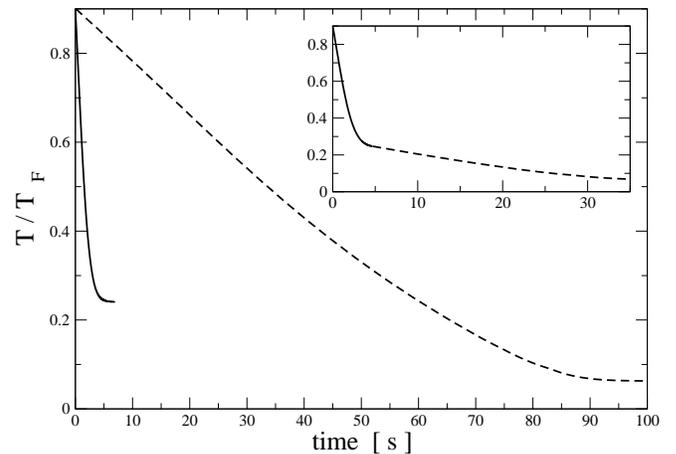}}
\caption{Temperature $T/T_F$ as a function of time for
$\gamma=10$ (solid line) and $\gamma=1.25$ (dashed line).
Here
$T_F=40.5\hbar\Omega_a$, and the observed $T_C\approx 0.6 T_F$.
The cooling is faster for $\gamma=10$, but the lower final temperature
of $0.06T_F$ is achieved for $\gamma=1.25$. The inset shows $T/T_F$
in a two stage cooling process where $\gamma$ is switched from $10$ to
$1.25$ at $T=0.25T_F$. }
\label{T_t}
\end{figure}

\begin{figure} 
\centering      
{\epsfig{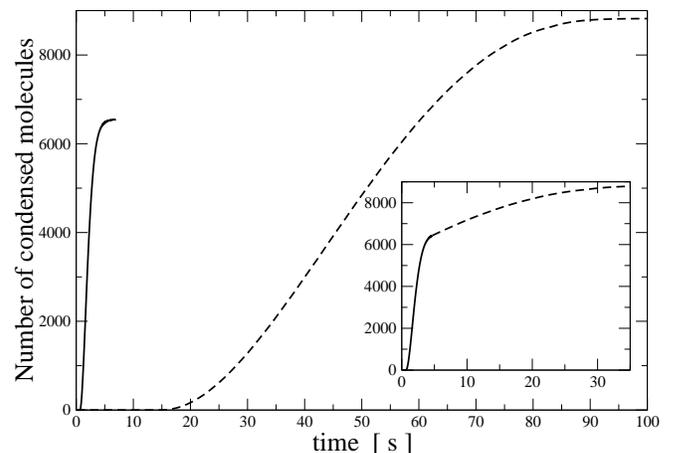}}
\caption{
The number of condensed molecules in the same simulations as in Fig. 1.
}
\label{Delta_t}
\end{figure}

In Fig. \ref{T_t} we show temperature as a function of time for laser
cooling with two values of $\gamma$. Similarly as on the BCS side of the
Feshbach resonance the nonzero gap function $\Delta(r)$ induced by the
molecular BEC does not affect much the efficiency of laser cooling.
Moreover, the cooling remains efficient even below the dissociation
temperature of $0.18T_F$. This is not quite surprising because even at
such low temperature there is a non-zero fraction of fermions induced by
coherent tunneling from the molecular BEC, which can be cooled by the
laser. The initial cooling is faster for $\gamma=10$ but the final
temperature of $0.06T_F$ is lower for $\gamma=1.25$. The best strategy to
reach low temperatures in reasonable time is to do the cooling in a few
stages starting with large $\gamma$ and finishing with small $\gamma$, as
shown in the inset in Fig.\ref{T_t}.  In Fig. \ref{Delta_t} we show the
growth of the molecular condensate in the same simulations as in
Fig.\ref{T_t}.

To summarize, our numerical simulations show that for reasonable
experimental parameters consistent with the requirements of the
Festina lente regime, it is possible to
laser cool fermions on the positive side of the Feshbach resonance
all the way down to the molecular BEC. Despite the pairing
effects, which produce a gap in the quasiparticle energy spectrum
and the heat capacity of non-condensed molecules, it is
possible to reach at least $0.06~T_F$ in a time of a few seconds.

                   
We would like to thank M. Baranov, L. Santos, and Z. Idziaszek
for discussions. This work was supported in part by ESF QUDEDIS program,
DFG (SFB 407, SPP1116, 432 POL), and the KBN grant PBZ-MIN-008/P03/2003.


\onecolumn
\section*{ Appendix A: Transition rates }

Laser cooling is realized by the Hamiltonian
\begin{equation}
\hat H_{las}=
\frac12\Omega
\int d^3r~ 
e^{i\vec k_{L}\vec r}
\hat\psi_{e}^{\dagger}(\vec r)
\hat\psi_{-}(\vec r)  ~+~
{\rm h.c.}~,
\end{equation}
driving coherent oscillations with Rabi frequency $\Omega$ between the
excited state $e$ and the state $-$, together with 
the spontaneous emission
superoperator
\begin{eqnarray}
{\cal L}~\hat\rho&=&
\gamma\int d\varphi~d\cos\theta~{\cal W}(\theta,\varphi)
\int d^3r_1~d^3r_2~
e^{ i \vec k ( \vec r_1 - \vec r_2 ) }
\nonumber\\
&\times&
\left[
2
\hat\psi_-^{\dagger}(\vec r_2)\hat\psi_e(\vec r_2)
\hat\rho~
\hat\psi_e^{\dagger}(\vec r_1)\hat\psi_-(\vec r_1)~-~
\hat\psi_e^{\dagger}(\vec r_1)\hat\psi_-(\vec r_1)
\hat\psi_-^{\dagger}(\vec r_2)\hat\psi_e(\vec r_2)
~\hat\rho~-~
~\hat\rho~
\hat\psi_e^{\dagger}(\vec r_1)\hat\psi_-(\vec r_1)
\hat\psi_-^{\dagger}(\vec r_2)\hat\psi_e(\vec r_2)
\right]~,
\end{eqnarray}
with an effective spontaneous emission rate of $2\gamma$
and the fluorescence dipole pattern ${\cal W}(\theta,\varphi)$.
Expansion in the eigenmodes $w_l(\vec r)$ of the harmonic trap
\begin{equation}
\hat\psi_e(\vec r)=\sum_l \hat e_l~w_l(\vec r)~,
\end{equation}
and the Bogoliubov transformation (see main text) give
\begin{equation}
\hat H_{las}=
\frac12\Omega
\sum_{ml}
\left[
\hat b_{m,-}           ~u_{ml}(\vec k_L)   ~+~
\hat b_{m,+}^{\dagger} ~v^*_{ml}(\vec k_L)
\right]~
\hat e_l^{\dagger}~+~
{\rm h.c.}~.
\end{equation}
Here the (generalized) Frank-Condon factors are
\begin{eqnarray}
&&
u_{ml}(\vec k)=
\int d^3r~
e^{i\vec k \vec r}
w_l^*(\vec r)
u_m(\vec r)~,
\nonumber\\
&&
v^*_{ml}(\vec k)=
\int d^3r~
e^{i\vec k \vec r}
w_l^*(\vec r)
v^*_m(\vec r)~.
\end{eqnarray}
In a similar way, and after rotating wave approximation for $e$-atoms
and the Bogoliubov quasiparticles, the spontaneous emission term becomes
\begin{eqnarray}
{\cal L}~\hat\rho&=&
\gamma \sum_{ml} U_{ml}
\left(
2\hat b_{m,-}^{\dagger}\hat e_l~\rho~\hat e_l^{\dagger}\hat b_{m,-} -
\hat e_l^{\dagger}\hat e_l \hat b_{m,-}\hat b_{m,-}^{\dagger}~\rho -
\rho~\hat e_l^{\dagger}\hat e_l \hat b_{m,-}\hat b_{m,-}^{\dagger}
\right)~+
\nonumber\\
&&
\gamma \sum_{ml} V_{ml}
\left(
2\hat b_{m,+}\hat e_l~\rho~\hat e_l^{\dagger}\hat b_{m,+}^{\dagger} -
\hat e_l^{\dagger}\hat e_l \hat b_{m,+}^{\dagger}\hat b_{m,+}~ \rho -
\rho~ \hat e_l^{\dagger}\hat e_l \hat b_{m,+}^{\dagger}\hat b_{m,+}
\right)~.
\end{eqnarray}
Here the spontaneous emission rates are
\begin{eqnarray}
&&
U_{ml}=
\int d\varphi d\cos\theta~  {\cal W}(\theta,\varphi)~
|u_{ml}(\vec k)|^2~,
\nonumber\\
&&
V_{ml}=
\int d\varphi d\cos\theta~  {\cal W}(\theta,\varphi)~
|v_{ml}(\vec k)|^2~.
\end{eqnarray}
Adiabatic elimination of the excited state $e$, similar as in
Ref.\cite{AE}, results in the following evolution equations for
the occupation numbers
\begin{eqnarray}
\frac{d N_{m,-}}{dt} &=&
\sum_n~
\Gamma^{(-)}_{m\leftarrow n} 
\left( 1 - N_{m,-} \right) N_{n,-} -
\Gamma^{(-)}_{n\leftarrow m} 
\left( 1 - N_{n,-} \right) N_{m,-} +
C_{mn} \left( 1 - N_{m,-} \right) \left( 1 - N_{n,+} \right) -
A_{nm} N_{m,-} N_{n,+} ~,
\\
\frac{d N_{m,+}}{dt} &=&
\sum_n~
\Gamma^{(+)}_{m\leftarrow n} 
\left( 1 - N_{m,+} \right) N_{n,+} -
\Gamma^{(+)}_{n\leftarrow m} 
\left( 1 - N_{n,+} \right) N_{m,+} +
C_{nm} \left( 1 - N_{m,+} \right) \left( 1 - N_{n,-} \right) -
A_{mn} N_{m,+} N_{n,-} ~.
\end{eqnarray}
There are contributions from four different processes:

\begin{itemize}

\item relaxation of $-$ quasiparticles 

\begin{equation}
\Gamma^{(-)}_{m\leftarrow n}=
\frac{\Omega^2}{2\gamma}
\sum_l
\frac{
\gamma^2 
U_{ml} 
|u_{nl}(\vec k_L)|^2
}
{
\left| 
\delta-\omega^e_l+
\omega_n+
i\gamma \sum_s~ 
U_{sl} ( 1 - N_{s,-} + \delta_{s,n} )+
V_{sl} N_{s,+}                    
\right|^2
}~,
\end{equation}

\item relaxation of $+$ quasiparticles

\begin{equation}
\Gamma^{(+)}_{m\leftarrow n}=
\frac{\Omega^2}{2\gamma}
\sum_l
\frac{
\gamma^2 
V_{nl} 
|v_{ml}(\vec k_L)|^2
}
{
\left| 
\delta-\omega^e_l-
\omega_m+
i\gamma \sum_s~ 
U_{sl} ( 1 - N_{s,-} )+
V_{sl} ( N_{s,+} + \delta_{m,s} )                    
\right|^2
}~,
\end{equation}

\item creation of pairs of $+$ and $-$ quasiparticles

\begin{equation}
C_{mn}=
\frac{\Omega^2}{2\gamma}
\sum_l
\frac{
\gamma^2 
U_{ml} 
|v_{nl}(\vec k_L)|^2
}
{
\left| 
\delta-\omega^e_l-
\omega_n+
i\gamma \sum_s~ 
U_{sl} ( 1 - N_{s,-} ) +
V_{sl} ( N_{s,+} + \delta_{n,s} )                    
\right|^2
}~,
\end{equation}

\item annihilation of pairs of $+$ and $-$ quasiparticles

\begin{equation}
A_{mn}=
\frac{\Omega^2}{2\gamma}
\sum_l
\frac{
\gamma^2 
V_{ml} 
|u_{nl}(\vec k_L)|^2
}
{
\left| 
\delta-\omega^e_l+
\omega_n+
i\gamma \sum_s~ 
U_{sl} ( 1 - N_{s,-} + \delta_{n,s} ) +
V_{sl} N_{s,+}                     
\right|^2
}~.
\end{equation}

\end{itemize}

\section*{ Appendix B: Spherical symmetry }

Simulation of laser cooling with a non-vanishing pairing function $P(\vec
r)$, and a Hartree potential $g_0\rho(\vec r)$ is a numerically hard 
problem. 
The
main difficulty is that after every short period of laser cooling it is
necessary to reiterate Bogoliubov-de Gennes equations in order to obtain a
new self-consistent pairing function, Hartree potential, and the
Bogoliubov modes $(\omega_m,u_m,v_m)$. With the new Bogoliubov modes, the
transition rates $A_{mn},C_{mn},\Gamma^{\pm}_{m\leftarrow n}$ have to be
calculated again,  and this is a very time-consuming operation. This
numerical effort is reduced a lot with assumption of spherical symmetry.
With spherically symmetric $\Delta(r)$ and $\rho(r)$ the Bogoliubov modes
can be decomposed into spherical and radial parts
\begin{eqnarray}
&&
u_{ln}(r)~Y_{lm}(\theta,\varphi)~,
\\
&&
v_{ln}(r)~Y_{lm}(\theta,\varphi)~.
\end{eqnarray}
In a similar way harmonic oscillator modes become
$W_{ln}(r)Y_{lm}(\theta,\varphi)$. Quasiparticle energies $\omega_{ln}$ do
not depend on $m$. We assume fast equilibration within each degenerate
energy shell $nl$ so that all occupation numbers in a given shell are the
same $N_{ln,\pm}$. This assumption greatly simplifies the rate equations
\begin{eqnarray}
\frac{d}{dt} N_{l_1n_1,-} 
&=\sum_{l_2n_2}&
\gamma^{(-)}_{l_1n_1\leftarrow l_2n_2}
\left( 1 - N_{l_1n_1,-} \right) N_{l_2n_2,-} -
\gamma^{(-)}_{l_2n_2\leftarrow l_1n_1}
\left( 1 - N_{l_2n_2,-} \right) N_{l_1n_1,-} +
\nonumber\\
&&
c_{l_1n_1l_2n_2} 
\left( 1 - N_{l_1n_1,-} \right) 
\left( 1 - N_{l_2n_2,+} \right) -
a_{l_2n_2l_1n_1} 
N_{l_1n_1,-} N_{l_2n_2,+} ~,
\\
\frac{d}{dt} N_{l_1n_1,+}
&=\sum_{l_2n_2}&
\gamma^{(+)}_{l_1n_1\leftarrow l_2n_2}
\left( 1 - N_{l_1n_1,+} \right) N_{l_2n_2,+} -
\gamma^{(+)}_{l_2n_2\leftarrow l_1n_1}
\left( 1 - N_{l_2n_2,+} \right) N_{l_1n_1,+} +
\nonumber\\
&&
c_{l_2n_2l_1n_1} 
\left( 1 - N_{l_1n_1,+} \right) 
\left( 1 - N_{l_2n_2,-} \right) -
a_{l_1n_1l_2n_2} N_{l_1n_1,+} N_{l_2n_2,-} ~.
\end{eqnarray}
Here $\gamma^{(\pm)},a,c$ are averaged transition rates, for example
\begin{equation}
\gamma^{(-)}_{l_1n_1\leftarrow l_2n_2}~=~
\frac{1}{2l_1+1}
\sum_{m_1m_2}
\Gamma^{(-)}_{l_1n_1m_1\leftarrow l_2n_2m_2}~.
\end{equation}
This equation combined together with the definition of $\Gamma^{(-)}$
gives
\begin{equation}
\gamma^{(-)}_{l_1n_1\leftarrow l_2n_2}~=~
\frac{\gamma\Omega^2}{2(2l_1+1)}
\sum_{m_1m_2}
\sum_{l_en_em_e}
\frac{
U_{l_1n_1m_1l_en_em_e}
|u_{l_2n_2m_2l_en_em_e}(\vec k_L)|^2
}
{
\left[ \delta-\omega(l_e+2n_e-\mu)+\omega_{l_2n_2}\right]^2+
\gamma^2 R_{l_en_e}^2
}~.
\end{equation}
Here the frequency of the intermediate excited state
$\omega(l_e+2n_e-\mu)$ is measured with respect to the chemical potential
$\mu$ because the quasiparticle energies $\omega_{l_2n_2}$ are also 
defined with respect to $\mu$. For a laser beam with a $\vec k_L$ 
along the $\hat z$-axis the indices $m_e$ and $m_2$ must be the same.
Furthermore, it follows from the properties of Clebsch-Gordan
coefficients that the sum $\sum_{m_1}U_{l_1n_1m_1l_en_em_e}$
does not depend on $m_e$ so that we can set e.g. $m_e=0$ and
\begin{eqnarray}
\gamma^{(-)}_{l_1n_1\leftarrow l_2n_2}&=&
\frac{\gamma\Omega^2}{2(2l_1+1)}
\sum_{l_en_e}
\frac{
\left(  \sum_{m_1} U_{l_1n_1m_1l_en_e0}  \right)
\left(  \sum_{m_2} |u_{l_2n_2m_2l_en_em_2}(\vec k_L)|^2  \right)
}
{
\left[ \delta-\omega(l_e+2n_e-\mu)+\omega_n \right]^2+
\gamma^2 R_{l_en_e}^2
}~\equiv
\nonumber\\
&&
\frac{\gamma\Omega^2}{2(2l_1+1)}
\sum_{l_en_e}
\frac{
SU_{l_1n_1l_en_e}~~  
su_{l_2n_2l_en_e}  
}
{
\left[ \delta-\omega(l_e+2n_e-\mu)+\omega_n \right]^2+
\gamma^2 R_{l_en_e}^2
}.
\end{eqnarray}
The last form shows that the transition rate $\gamma^{(-)}$ can be
constructed out of matrices $SU,su$ and a vector $R$. These
elements can be expressed through even more elementary building blocks
\begin{eqnarray}
&&
SU_{l_1n_1l_en_e}~=~
\sum_l
(jWu)^2_{ll_en_el_1n_1}~
S_{ll_el_1}~,
\nonumber\\
&&
su_{l_2n_2l_en_e}~=~
\sum_m
\left|
\sum_l
s_{ll_el_2m}
(jWu)_{ll_en_el_2n_2}
\right|^2~.
\nonumber
\end{eqnarray}
The more elementary building blocks are
\begin{eqnarray}
&&
(jWu)_{ll_en_el_1n_1}=
\int r^2dr~j_l(k r)W_{l_en_e}(r)u_{l_1n_1}(r)~,
\nonumber\\
&&
S_{ll_el_1}=
(2l+1)
\frac{2l_e+1}{2l_1+1}
\sum_m
\langle l,l_e,m,0 | l_1,m \rangle^2
\langle l,l_e,0,0 | l_1,0 \rangle^2~,
\nonumber\\
&&
s_{ll_el_2m}=
i^l(2l+1)
\sqrt{\frac{2l_e+1}{2l_2+1}}
\langle l,l_e,0,m|l_2,m \rangle
\langle l,l_e,0,0|l_2,0 \rangle~.
\end{eqnarray}
The line-width of an excited state $l_e,n_e$ is approximately given by
\begin{equation}
R_{l_en_e}=
\sum_{ln}
(1-N_{ln})
SU_{lnl_en_e} +
N_{ln}
SV_{lnl_en_e}~.
\end{equation}
The matrix $SV$ is obtained from the matrix $SU$
by a substitution $u\to v$. As the 
eigenfunction $u_{ln}(r)$ evolves in the process of laser cooling
it is more efficient to express the matrix $(jWu)$ through a static
matrix $(jWW)$ build out of harmonic oscillator modes
\begin{equation}
(jWW)_{ll_en_el_1n_1}=
\int r^2dr~j_l(k r)W_{l_en_e}(r)W_{l_1n_1}(r)~.
\end{equation}
The static matrices $s,S,jWW$ were prepared once,  and stored then on a 
disk.
They facilitate calculation of the temperature dependent matrices
$SU,su,R$ every time the wave functions $u,v$ are updated.

\end{document}